\documentstyle[12pt]{article}

\begin{document}

\title{The Supersymmetry of  Relativistic P\" oschl-Teller  Systems}

\author{Ion I. Cot\u aescu\\ {\small \it The West 
University of Timi\c soara,}\\ {\small \it V. P\^ arvan Ave. 4, RO-1900 
Timi\c soara, Romania}}

\date{\today}

\maketitle

\begin{abstract}
We analyze the supersymmetry and the shape invariance of the  potentials of the 
(1+1) relativistic oscillators we have recently proposed \cite{COTA,COTA1}.

\end{abstract}

In the general relativity, the geometric models play the role of  
kinematics, helping us to understand the characteristics of the  classical or 
quantum free motion on a given background. One of the simplest (1+1) geometric 
models is that of the  quantum relativistic harmonic oscillator   represented 
by a free massive scalar field  on the  anti-de Sitter  static background 
\cite{OSC,M}.  
Recently,  we have generalized this model to a  family of quantum models the 
metrics of which depend on a real parameter \cite{COTA}. We have shown 
\cite{COTA1} that this family contains a set of models which can be considered 
as the relativistic correspondents of the usual  nonrelativistic 
P\" oschl-Teller (PT) systems \cite{PT1,PT2}. These models have been referred 
as relativistic PT systems \cite{COTA1}. They have countable discrete energy 
spectra and the same square integrable energy  eigenfunctions as those known 
from the  nonrelativistic quantum mechanics. However, the significance of the 
parameters as well as the formula of the energy levels are different.  An 
important property of all these models is that  they lead to  the  usual 
harmonic oscillator in the nonrelativistic limit.

These analytically solvable relativistic problems may have similar properties 
of supersymmetry and shape invariance as those known from the  nonrelativistic 
theory \cite {PT3}. Here, our aim is to analyze these properties, pointing out 
the changes requested by the specific form of the Klein-Gordon operator and by 
the relativistic expression of the energy levels.

Let us start with a short review of the main results concerning the 
relativistic 
PT problem (in natural units, $\hbar=c=1$) \cite{COTA1}. There exists an 
holonomic frame, $(t,x)$, in which the metrics of our family of models is given 
by the line element 
\begin{equation}
ds^{2}=\left(1+\frac{1}{\epsilon^2}\tan^{2}\hat\omega x\right)(dt^{2}-
dx^{2}).
\end{equation} 
where we have denoted $\hat\omega=\epsilon\omega$ understanding that $\omega$ 
isthe frequency of the harmonic oscillator obtained as the nonrelativistic 
limit of these models. Here $\epsilon$ is a new parameter which determines  
the domain of the quantum free motion, $D =(-\pi/2\hat\omega, \pi/2\hat\omega)$, 
for a fixed $\omega$.  Note that for $\epsilon=1$ we obtain  the exact anti-de 
Sitter metric. Particularly, in the frame $(t,x)$, the scalar product of the 
square integrable functions defined on $D$ is just the usual one. The 
Klein-Gordon equation of a spinless particle of the mass $m$ has the form
\begin{equation}\label{(kg2)}
\left(-\frac{d^2}{dx^{2}} +\frac{m^2}{\epsilon^2}\tan^{2}\hat\omega x
\right)U(x)=({E_{n}}^{2}-m^{2})U(x).
\end{equation}
It has a countable discrete energy spectrum. The corresponding square- 
integrable energy eigenfunctions  are
\begin{equation}\label{(U2)}
U_{n_{s},s}(x)=N_{n_{s},s} \cos^{k}\hat\omega x 
\sin^{s} \hat\omega x F(-n_{s}, k+s+n_{s},
s+\frac{1}{2},\sin^{2} \hat\omega x),
\end{equation}
where $N_{n_{s},s}$ is the normalization factor and
\begin{equation}\label{(p)}
k=\frac{1}{2}\left[1 + \sqrt{1+ 4\frac{m^{2}}{\epsilon^{2}\hat\omega^{2}}}
\right] > 1
\end{equation}\label{(wei)}
is the positive solution of the equation
\begin{equation}\label{(kk)}
k(k-1)=\frac{m^2}{\epsilon^{2}\hat\omega^2}.
\end{equation}
The quantum numbers, $n_{s}=0,1,2...$ and $s=0, 1$, 
can be embedded into the main quantum number $n=2n_{s}+s$. This will take 
even values if $s=0$ and odd values for $s=1$. Hence, the functions 
$U_{n_{s},s}(x)$ are completely determined by  $n$ only. They 
are real polynomials of the degree $n$ in $\sin\hat\omega x$, with the factor 
$\cos^{k}\hat\omega x$, which vanishes at  $x=\pm \pi/2\hat\omega$.  
The energy levels also depend  on the main quantum number. These can be 
obtained from the quantization condition 
\begin{equation}\label{(el1)}
{E_{n}}^{2}-m^{2}\left(1-\frac{1}{\epsilon^{2}}\right)=\hat\omega^{2}(n+k)^{2}, \quad 
n=0,1,2... .
\end{equation}  
We specify that, for the pure anti-de Sitter oscillator, the resulting energy 
spectrum is equidistant \cite{M} since $\epsilon=1$. 

We have seen that the relativistic PT systems depend only on three parameters, 
the mass $m$,  the frequency $\omega$  and the new parameter $\epsilon$. For 
our future needs it is convenient to change this parametrization by using the 
parameter $k$ instead of $m$. Then, from (\ref{(kk)}) we have
\begin{equation}\label{(mm)}
m^2=k(k-1)\epsilon^{2}\hat\omega^{2},
\end{equation}
with the help of which the quantization condition (\ref{(el1)}) gives
\begin{equation}\label{(el2)}
{E_{n}}^2=\hat\omega^{2}[(n+k)^{2}+(\epsilon^{2}-1)k(k-1)].
\end{equation} 
Moreover, the second term of the left-hand side of Eq.(\ref{(kg2)}), which has 
been called  the relativistic PT potential \cite{COTA1}, can be written now as   
\begin{equation}
V_{PT}(k,x)=k(k-1)\hat\omega^2\tan^{2}\hat\omega x.
\end{equation}
In the following we shall consider that the  parameters $\omega$ and $\epsilon$ 
remain fixed, while $k$ is  variable. For this reason, and by taking into account 
that the eigenfunctions (\ref{(U2)}) are defined, in fact, only by the main 
quantum number, we shall denote $U_{n_{s},s}\equiv U_{k,n}$.

In order to construct the supersymmetric quantum mechanics \cite{PT3} of the  
relativistic PT systems we have to  adjust the  Klein-Gordon operator 
(\ref{(kg2)}) so that its spectrum is 
\begin{equation}
({E_{n}}^{2}-{E_{0}}^{2})/\hat\omega^{2}=n(n+2k).
\end{equation}
Thus, we obtain the new  operator $\Delta[V]$ defined by  
\begin{equation}
\{\Delta[V]U\}(x)=\left(-\frac{1}{\hat\omega^2}\frac{d^2}{dx^2}+V(x)\right)U(x).
\end{equation}
which will play here the same role as the Hamiltonian of the nonrelativistic 
theory. With its help  we can rewrite the Klein-Gordon equation  in the  form 
\begin{equation}\label{(o1)}
\Delta[V_{-}(k)]U_{k,n}=n(n+2k)U_{k,n}
\end{equation}
where we have denoted
\begin{equation}
V_{-}(k,x)=\frac{1}{\hat\omega^2}[V_{PT}(k,x)+m^{2}-E_{0}^2]=k(k-1)
\tan^{2}\hat\omega x -k.
\end{equation}
Now, from the (normalized \cite{B2}) ground-state eigenfunction 
\begin{equation}\label{(eg)}
U_{k,0}(x)=\left(\frac{\hat\omega^2}{\pi}\right)^{\frac{1}{4}}\left[
\frac{\Gamma(k+1)}{\Gamma(k+\frac{1}{2})}\right]^{\frac{1}{2}}
\cos^{k}\hat\omega x.
\end{equation}
we obtain the superpotential
\begin{equation}
W(k,x)=-\frac{1}{\hat\omega}\frac{1}{U_{k,0}(x)}\frac{dU_{k,0}(x)}{dx}=
k\tan\hat\omega x
\end{equation} 
and the supersymmetric partner of $V_{-}$
\begin{equation}
V_{+}(k,x)=-V_{-}(k,x)+2W(k,x)^{2}=k(k+1)\tan^{2}\hat\omega x +k.
\end{equation}
Furthermore, following the standard procedure \cite{PT3},  we can define the 
pair of the lowering and raising operators, $A_{k}$ and $A_{k}^{+}$, having 
the action 
\begin{eqnarray}
(A_{k}U)(x)&=&\left(\frac{1}{\hat\omega}\frac{d}{dx}+W(k,x)\right)U(x),\\ 
(A_{k}^{+}U)(x)&=&\left(-\frac{1}{\hat\omega}\frac{d}{dx}+W(k,x)\right)U(x).
\end{eqnarray}
and the commutation rule
\begin{equation}
[A_{k},A_{k}^{+}]=2k+\frac{1}{2k}(A_{k}+A_{k}^{+})^2.
\end{equation}
Then, the operators $\Delta$ of the systems with superpartner potentials, 
$V_{-}$ and $V_{+}$, can be written as      
\begin{equation}
\Delta[V_{-}(k)]=A_{k}^{+}A_{k} , \qquad
\Delta[V_{+}(k)]=A_{k}A_{k}^{+}.
\end{equation}
Let us observe now that these potentials are shape invariant since
\begin{equation}
V_{+}(k,x)=V_{-}(k+1,x)+2k+1.
\end{equation}
Consequently, we can verify that 
\begin{equation}\label{(o2)}
\Delta[V_{+}(k)]U_{k+1,n-1}=n(n+2k)U_{k+1,n-1}.
\end{equation}
From this equation  combined with  (\ref{(o1)}) it results that the normalized 
energy eigenfunctions satisfy
\begin{equation}\label{(aa)}
A_{k}U_{k,n}=\sqrt{n(n+2k)}U_{k+1,n-1}, \qquad
A_{k}^{+}U_{k+1,n-1}=\sqrt{n(n+2k)}U_{k,n}. 
\end{equation}

Hence, we have obtained the relation between a pair of models with  
superpartner potentials. These  can correspond to any pair of values
$k$ and $k+1$ of our variable  parameter. Therefore, if we consider the 
sequence of models with $k=k_{0}, k_{0}+1, 
k_{0}+2,..$ ($k_{0}>1$), then the of successive models are  
supersymmetric partners. What is interesting here is that the masses of these 
models appear as being quantized because of Eq. (\ref{(mm)}) where $k$ play the 
role of the quantum number. On the other hand, we observe that the action 
(\ref{(aa)}) of the operators $A_{k}$ and $A_{k}^{+}$ leaves invariant the 
value of $n+k$. This indicates that the operator $H$, defined by 
$HU_{n,k}=(n+k)U_{n,k}$,  will  commute with these operators. Finally we note 
that the very simple form of the shape invariance here allows one to write any 
normalized energy eigenfunction as
\begin{equation}
U_{k,n}=\frac{1}{\sqrt{n!}}\left[\frac{\Gamma(n+2k)}{\Gamma(2n+2k)}
\right]^{\frac{1}{2}}
A_{k}^{+}A_{k+1}^{+}...A_{k+n-1}^{+}U_{k+n,0}.
\end{equation}
where $U_{k+n,0}$ is the normalized ground-state eigenfunction given by 
(\ref{(eg)}).

As a conclusion,  we can say that our example of relativistic supersymmetric 
quantum mechanics of a massive spinless particle follows the same general rules 
as in 
the nonrelativistic case. The unique major difference is that  the Hamiltonian 
of the nonrelativistic theory is replaced here by the operator $\Delta$ which 
is linearly dependent on the squared Hamiltonian. For this reason, the 
commutation relations of the Hamiltonian with the raising and lowering 
operators will be  different too. In our opinion this could lead to new 
algebraic properties of the relativistic PT systems.

\end{document}